\documentstyle[12pt]{article}
\begin{document}

\def\p{\phi}
\def\P{\Phi}
\def\a{\alpha}
\def\e{\varepsilon}
\def\be{\begin{equation}}
\def\ee{\end{equation}}
\def\l{\label}
\def\0{\setcounter{equation}{0}}
\def\T{\hat{T}_}
\def\b{\beta}
\def\S{\Sigma}
\def\3{d^3{\rm \bf x}}
\def\4{d^4}
\def\C{\cite}
\def\r{\ref}
\def\ba{\begin{eqnarray}}
\def\ea{\end{eqnarray}}
\def\n{\nonumber}
\def\R{\right}
\def\L{\left}
\def\q{\hat{Q}_0}
\def\X{\Xi}
\def\x{\xi}
\def\la{\lambda}
\def\d{\delta}
\def\s{\sigma}
\def\f{\frac}
\def\vx{{\rm \bf x}}
\def\j{\frac{\delta}{i \delta j_a ({\rm \bf x},x_0+t+t_1)}}

\begin{titlepage}
\begin{flushright}
{\normalsize IP GAS-HE-8/95}
\end{flushright}
\vskip 3cm
\begin{center}
{\Large\bf On the cancelation of quantum-mechanical corrections at the
periodic motion}
\vskip 1cm

\mbox{J.Manjavidze}\footnote{Institute of Physics,
Georgian Academy of Sciences, Tamarashvili str. 6,
Tbilisi 380077, Republic of Georgia,
e-mail:~jm@physics.iberiapac.ge} \\

\end{center}
\date{MARCH 1995}
\vskip 1.5cm

\begin{abstract}
\footnotesize
The paper contains description of the path integrals in the
action-angle phase space. It allows to split the action and angle
quantum degrees of freedom and to show that the angular quantum
corrections are cancel each other if the classical trajectory
is periodic. The considered in the paper example shows that the
quantum problem can be quasiclassical over the part (angular in
the considered case) degrees of freedom.

\end{abstract}

\end{titlepage}

\section{Introduction}
\setcounter{equation}{0}

It will be shown in this paper that the quantum fluctuations of
angular variables cancel each other if the classical motion is periodic.
This cancelation mechanism can be used for the path-integral explanation
of the rigid rotator problems integrability (last one is isomorphic
to the Pocshle-Teller problem \C {pt}) \C {duru}. Note also that the classical
trajectories of all known integrable quantum-mechanical problems (of
the rigid rotator, of the H-atom \C {col}, etc.) are periodic.

Our technical problem consist in necessity to extract the
quantum angular degrees of freedom. For this purpose  we will
use the unitary definition of the path integral measure which
guarantees the conservation of total probability at arbitrary
transformations of the path integral variables \C{manj}. It will
allow to define the path integral in the phase space of
action-angle variables and, correspondingly, to define the quantum
measure of the angular degrees of freedom.

Mostly probable that the considered phenomena has the general
character and its demonstration will be fruitful. For
simplicity this effect of cancellations we will demonstrate
on the one-dimensional $\lambda x^4$ model \C{tich}. In the following
section the brief description of unitary definition of the
path-integral measure will be given. The perturbation theory in
terms of action-angle variables will be constracted in Sec.3
(the scheme of transformed perturbation theory was given in
\C{manj}). In Sec.4 the cancelation mechanism will be demonstrated.

\section{The unitary definition of the path-integral measure}
\setcounter{equation}{0}

We will calculate the the probability
\be
R(E)= \int dx_1 dx_2 |A(x_1 ,x_2 ;E )|^2 ,
\l{1}
\ee
to introduce the unitary definition of  path-integral measure
\C{yp}. Here
\be
A(x_1 ,x_2 ;E )=i\int^{\infty}_{0} dT e^{iET}
\int_{x(0)=x_1}^{x(T)=x_2} Dx e^{iS_{C_+ (T)}(x)}
\l{2}
\ee
is the  amplitude. The action
\be
S_{C_+ (T)}(x)=
\int_{C_+ (T)}  dt (\frac{1}{2}\dot{x}^2 -
\frac{\omega_{0}^{2}}{2}x^2-\frac{\lambda}{4}x^4)
\l{3}
\ee
is defined on the Mills' contour \C{mil}:
\be
C_{\pm} (T): t\rightarrow t\pm i\epsilon,
\;\;\;\epsilon\rightarrow +0,
\;\;\;0\leq t\leq T.
\l{4}
\ee
So, we will omit the calculation of the amplitude since it is sufficient
to now $R(E)$ for the bound states energies computation (see also \C{ann}
where a many-particles system was considered from this point of view).

Inserting (\ref{2}) into (\ref{1}) we find:
\be
R(E)=
\int^{\infty}_{0} dT_+ dT_-  e^{iE(T_+ -T_- )}
\int^{x_+(T_+)=x_-(T_-)}_{x_+(0)=x_-(0_-)} Dx e^{iS_{C^{-}} (x)}
\l{5}
\ee
is described by the closed-path integral. The  total action
\be
S_{C^{-}} (x)=
S_{C_+ (T_+ )} (x_+ )- S_{C_{-}(T_- )} (x_- ),
\l{6}
\ee
where the integration over turning points
\be
x_1 =x_+ (0)=x_- (0),\;\;\; x_2 =x_+ (T_+ )=x_- (T_- )
\l{8}
\ee
was performed.

Using the linear transformations:
\be
x_{\pm}(t)=x(t) \pm e(t)
\l{9}
\ee
and
\be
T_{\pm}=T \pm \tau
\l{10}
\ee
we find, calculating integrals over $e(t)$ and $\tau$ perturbatively
\C{manj}, that
\ba
R(E)=2\pi \int^{\infty}_{0} dT
\exp\{\frac{1}{2i}\hat{\omega}\hat{\tau}-i
\int_{C^{(+)}(T)} dt
\hat{j}(t)\hat{e}(t)\}
\int Dx e^{-i\tilde{H}(x;\tau )-iV_T (x,e)}\times
\n \\ \times
\delta (E+ \omega -H_T (x))
\prod_{t} \delta (\ddot{x} +\omega_{0}^{2}x +\lambda x^3 -j ).
\l{11}
\ea
The ``hat" symbol means differentiation over corresponding auxiliary
quantity. For instance,
\be
\hat{\omega}\equiv\frac{\partial}{\partial \omega}, \;\;\;
\hat{j}(t)=\frac{\delta}{\delta j(t)}.
\l{12}
\ee
It will be assumed that
\ba
\hat{j}(t\in C_{\pm})j(t'\in C_{\pm})=\delta (t-t'),
\n \\
\hat{j}(t\in C_{\pm})j(t'\in C_{\mp}) \equiv 0.
\l{i}
\ea
The time integral over contour $C^{(\pm)}(T)$ means that
\be
\int_{C^{(\pm)}(T)} =\int_{C_{+}(T)} \pm \int_{C_{-}(T)}.
\l{14}
\ee
At the end of calculations the limit $(\omega ,\tau, j, e)=0$
must be calculated. The explicit  form of $\tilde{H}(x;\tau )$,
$V_T (x,e)$ will be given later; $H_T (x)$ is the Haniltonian
at the  time moment $t=T$.

The functional $\delta$-function unambiguously determines the
contributions in the path integral. For this purpose we must
find the strict solution $x_j (t)$ of the equation of motion:
\be
\ddot{x} +\omega_{0}^{2}x +\lambda x^3 -j =0,
\l{15}
\ee
expanding it over $j$. In zero order over $j$ we have
the classical trajectory $x_c$ which is defined by the equation of
motion:
\be
\ddot{x} +\omega_{0}^{2}x +\lambda x^3  =0.
\l{16}
\ee
This equation is equivalent to the following one:
\be
t+\theta_0 =\int^{x} dx \{2(h-\omega_{0}^{2}x^2 -\lambda x^4 ) \}^{-1/2}.
\l{17}
\ee
The solution of this equation is the periodic elliptic function
\C{tb}. Here $(h, \theta_0)$ are the constants of integration of
eq.(\ref{16}).

The mapping of our problem on the action-angle phase space will be
performed using representation (\ref{11}) \C{manj}. Using the
obvious definition of the  action:
\be
I=\frac{1}{2\pi}\oint \{2(h-\omega_{0}^{2}x^2 -\lambda x^4 ) \}^{1/2},
\l{18}
\ee
and of the angle
\be
\phi =\frac{\partial h}{\partial I}
\int^{x_c } \{2(h-\omega_{0}^{2}x^2 -\lambda x^4 ) \}^{-1/2}
\l{19}
\ee
variables \C{arn} we easily find from (\ref{11}) that
\ba
R(E)=2\pi \int^{\infty}_{0} dT
\exp\{\frac{1}{2i}\hat{\omega}\hat{\tau}-i
\int_{C^{(+)}(T)} dt \hat{j}(t)\hat{e}(t)\}
\int DI D\phi e^{-i\tilde{H}(x_c ;\tau )-iV_T (x_c ,e)}\times
\n \\ \times
\delta (E+ \omega -h_T (I))
\prod_{t} \delta (\dot{I} -j\frac{\partial x_c}{\partial \phi})
\delta (\dot{\phi} -\Omega (I) +j\frac{\partial x_c}{\partial I}),
\l{20}
\ea
where $x_c =x_c (I,\phi)$ is the solution of eq.(\ref{19}) with
$h=h(I)$ as the solution of eq.(\ref{18}) and the  frequency
\be
\Omega (I)=\frac{\partial h}{\partial I}.
\l{21}
\ee
Representation (\ref{20}) is not the full solution of our problem:
the action and angle variables are still interdependent since they
both are exited by the  same source $j(t)$. This reflects the Lagrange
nature of the path-integral description of $(x,p)$ phase-space motion. The
true Hamiltone's description must contain independent quantum sources
of action and angle variables.

\section{The perturbation theory in the action-angle phase
space}
\setcounter{equation}{0}

The structure of source terms $j\partial x_c/\partial \phi$
and $j\partial x_c/\partial I$ shows that the source of quantum
fluctuations is the classical trajectories perturbations and $j$ is the
auxiliary variable. It allows to regroup the perturbation series  in a
following manner. Let us consider the action of the
perturbation-generating operators:
\ba
e^{-i\int_{C^{(+)}(T)} dt \hat{j}(t)\hat{e}(t) }
e^{-iV_T (x,e)}
\prod_{t} \delta (\dot{I} +j\frac{\partial x_c}{\partial \phi})
\delta (\dot{\phi} -\Omega (I) -j\frac{\partial x_c}{\partial I}=
\n \\=
\int De_I De_{\phi}
e^{i\int_{C^{(+)}}dt (e_I \dot{I}+e_{\phi}(\dot{\phi}-\Omega (I)))}
e^{-iV_T (x,e_c )},
\l{22}
\ea
where
\be
e_c (e_I ,e_{\phi})=
e_I \frac{\partial x_c}{\partial \phi} -
e_{\phi} \frac{\partial x_c}{\partial I}
\l{23}
\ee
The integrals  over $(e_I ,e_{\phi})$ will be calculated perturbatively:
\be
e^{-iV_T (x,e_c )}=
\sum^{\infty}_{n_I ,n_{\phi} =0}
\frac{1}{n_I !n_{\phi} !}
\int \prod^{n_I }_{k=1}(dt_k e_I (t_k))
\prod^{n_{\phi} }_{k=1}(dt'_k e_{\phi} (t'_k))
P_{n_I ,n_{\phi}} (x_c ,t_1 ,...,t_{n_I},t'_1,...,t_{n_{\phi}}),
\l{24}
\ee
where
\be
P_{n_I ,n_{\phi}} (x_c ,t_1 ,...,t_{n_I},t'_1,...,t_{n_{\phi}})=
\prod^{n_I }_{k=1} \hat{e}'_I (t_k)
\prod^{n_{\phi} }_{k=1} \hat{e}'_{\phi} (t'_k)
e^{-iV_T (x,e'_c )}.
\l{25}
\ee
Here $e'_c \equiv e_c (e'_I ,e'_{\phi} )$ and the derivatives in this
equality are calculated at $e'_I =0$, $e'_{\phi}=0$. At the same time,
\be
\prod^{n_I }_{k=1} e_I (t_k)
\prod^{n_{\phi} }_{k=1} e_{\phi} (t'_k)=
\prod^{n_I }_{k=1} (i\hat{j}_I (t_k))
\prod^{n_{\phi} }_{k=1} (i\hat{j}_{\phi} (t'_k))
e^{-i\int_{C^{(+)}} dt (j_I (t)e_I (t)+j_{\phi}(t)e_{\phi}(t))}.
\l{26}
\ee
The limit $(j_I , j_{\phi})=0$ is assumed. Inserting (\ref{24}),
(\ref{26}) into (\ref{22}) we will find new representation for $R(E)$:
\ba
R(E)=2\pi \int^{\infty}_{0} dT
\exp\{\frac{1}{2i}\hat{\omega}\hat{\tau}-i
\int_{C^{(+)}(T)} dt (\hat{j}_I (t)\hat{e}_I (t)+
\hat{j}_{\phi} (t)\hat{e}_{\phi} (t) )\}\times
\n \\ \times
\int DI D\phi e^{-i\tilde{H}(x_c ;\tau )-iV_T (x_c ,e_c )}
\delta (E+ \omega -h_T (I))
\prod_{t} \delta (\dot{I} -j_I )
\delta (\dot{\phi} -\Omega (I) -j_{\phi})
\l{27}
\ea
in which the action and the angle degrees of freedom are decoupled.

Solving the canonical equations of motion:
\be
\dot{I} =j_I,\;\;\;\dot{\phi} =\Omega (I) +j_{\phi},
\l{28}
\ee
the boundary conditions:
\be
I_j (0)=I_0,\;\;\;\phi_j (0)=\phi_0
\l{29}
\ee
for the solutions $I_j ,\phi_j $ of eqs.(\ref{28}) will be used.
This will lead to the following Green function:
\be
g(t-t')=\Theta (t-t'),
\l{30}
\ee
with symmetric step function: $\Theta  (0)=1/2$.
The solutions of eqs.(\ref{28})  have the  form:
\ba
I_j (t)=I_0 +
\int dt' g(t-t') j_I (t')
\equiv I_0 +I'(t),
\n \\
\phi_j (t)=\phi_0 + \tilde{\Omega}(I_j ) t+
\int dt' g(t-t') j_{\phi} (t') \equiv
\phi_0 +\tilde{\Omega}(I_j )t+\phi' (t),
\l{31}
\ea
where
\be
\tilde{\Omega}(I_j ) =\frac{1}{t}\int dt' g(t-t') \Omega (I_0 +I'(t')).
\l{32}
\ee
Inserting (\ref{31}) into (\ref{27}) we find:
\ba
R(E)=2\pi \int^{\infty}_{0} dT\exp\{\frac{1}{2i}\hat{\omega}\hat{\tau}-i
\int_{C^{(+)}(T)} dt (\hat{j}{_I}(t)\hat{e}_I (t)+
\hat{j}_{\phi}(t)\hat{e}_{\phi} (t))\}\times
\n \\ \times
\int^{\infty}_{0} dI_0 \int^{2\pi}_{0} d\phi_0
e^{-i\tilde{H}(x_c ;\tau )-iV_T (x_c ,e_c )}
\delta (E+ \omega -h_T (I_j )),
\l{33}
\ea
where
\be
x_c =x_c (I_j ,\phi_j )=
x_c (I_0 +I'(t) ,\phi_0 +\tilde{\Omega}(I_j )t+\phi' (t) )
\l{34}
\ee
and $e_c$ was defined in (\ref{23}). Note that the
measure of the integrals over $(I_0 ,\phi_0 )$ was  defined without of
the Faddeev-Popov's ansatz \C {fp} and there is not any ``hosts"
since the Jacobian of transformation is equal to one.

We can extract the Green function into the  perturbation-generating
operator using the equalities:
\ba
\hat{j}_I (t)=\int dt' g(t-t') \hat{I}'(t),
\n \\
\hat{j}_{\phi}=\int dt' g(t-t') \hat{\phi}'(t),
\l{35}
\ea
which evidently follows from (\ref{31}). In result,
\ba
R(E)=2\pi \int^{\infty}_{0} dT\exp\{\frac{1}{2i}\hat{\omega}\hat{\tau}
-i\int_{C^{(+)}(T)} dtdt' g(t'-t) (\hat{I}(t)\hat{e}_I (t')+
\hat{\phi}(t)\hat{e}_{\phi} (t'))\}\times
\n \\ \times
\int^{\infty}_{0} dI_0 \int^{2\pi}_{0} d\phi_0
e^{-i\tilde{H}(x_c ;\tau )-iV_T (x_c ,e_c )}\times
\n \\ \times
\delta (E+ \omega -h_T (I_j )),
\l{36}
\ea
where $x_c$ was defined  in (\ref{34}).

We can define the formalism without doubling of degrees of
freedom. One can use the fact that the action of perturbation-generating
operators and the analytical continuation to the real times are the
commuting operations. This  can be seen easily using the definition
(\ref{i}). In result:
\ba
R(E)=2\pi \int^{\infty}_{0} dT\exp\{\frac{1}{2i}\hat{\omega}\hat{\tau}-i
\int_0^T dtdt' \Theta (t'-t) (\hat{I}(t)\hat{e}_I (t')+
\hat{\phi}(t)\hat{e}_{\phi} (t'))\}\times
\n \\ \times
\int^{\infty}_{0} dI_0 \int^{2\pi}_{0} d\phi_0
e^{-i\tilde{H}(x_c ;\tau )-iV_T (x_c ,e_c )}
\delta (E+ \omega -h_T (I_0+ I(T)),
\l{37}
\ea
where
\be
\tilde{H}_T (x_c ;\tau )=
2\sum^{\infty}_{n=1}\frac{\tau^{2n+1}}{(2n+1)!}
\frac{d^{2n}}{dT^{2n}}h(I_0 +I(T))
\l{38}
\ee
and
\be
-V_T (x_c ,e_c )=S(x_c +e_c )-S(x_c -e_c ).
\l{39}
\ee
Now we will use the  last $\delta$-function:
\ba
R(E)=2\pi \int^{\infty}_{0} dT\exp\{\frac{1}{2i}(\hat{\omega}\hat{\tau}+
\int_0^T dtdt' \Theta (t'-t) (\hat{I}(t)\hat{e}_I (t')+
\hat{\phi}(t)\hat{e}_{\phi} (t')) \}\times
\n \\ \times
\int^{\infty}_{0} dI_0 \int^{2\pi}_{0}
\frac{d\phi_0}{\Omega (E+\omega)}
e^{-i\tilde{H}(x_c ;\tau )-iV_T (x_c ,e_c )},
\l{40}
\ea
Here
\be
x_c (t)=x_c (I(E+\omega )+I(t)-I(T), \phi_0 +\tilde{\Omega}t+\phi (t)).
\l{41}
\ee

Eq.(\ref{40}) contains unnecessary contributions: the  action
of the operator
\be
\int^{T}_{0}dt dt' \Theta (t-t') \hat{e}_I (t)\hat{I}(t')
\l{42}
\ee
on $\tilde{H}_T$, defined in (\ref{38}), leads to the time integrals
with zero integration range:
\be
\int^{T}_{0}dt \Theta (T-t) \Theta (t-T) =0.
\l{43}
\ee
Using this fact,
\ba
R(E)=2\pi \int^{\infty}_{0} dT  e^{\frac{1}{2i}\int_0^T dtdt' \Theta (t'-t)
(\hat{I}(t)\hat{e}_I (t')+
\hat{\phi}(t)\hat{e}_{\phi} (t'))}\times
\n \\ \times
\int^{\infty}_{0} dI_0 \int^{2\pi}_{0} \frac{d\phi_0}{\Omega (E)}
e^{-iV_T (x_c ,e_c )},
\l{44}
\ea
where
\be
x_c (t)=x_c (I_0 (E)+I(t)-I(T), \phi_0 +\tilde{\Omega}t+\phi (t)).
\l{45}
\ee
is the periodic function:
\be
x_c (I_0 (E)+I(t)-I(T), (\phi_0 +2\pi ) +\tilde{\Omega}t+\phi (t))=
x_c (I_0 (E)+I(t)-I(T), \phi_0 +\tilde{\Omega}t+\phi (t)).
\l{ii}
\ee
Now we can consider the cancelation of angular perturbations.

\section{Cancelation of angular perturbations}
\setcounter{equation}{0}

Introducing the perturbation-generating operator into the integral
over $\phi_0$:
\ba
R(E)=2\pi \int^{\infty}_{0} dT  e^{\frac{1}{2i}
\int_0^T dtdt' \Theta (t'-t) \hat{I}(t)\hat{e}_I (t')}\times
\n \\ \times
\int^{\infty}_{0} dI_0 \int^{2\pi}_{0} \frac{d\phi_0}{\Omega (E)}
e^{\frac{1}{2i}\int_0^T dtdt'
\Theta (t'-t) \hat{\phi}(t)\hat{e}_{\phi} (t')} e^{-iV_T (x_c ,e_c )},
\l{46}
\ea
the mechanism of cancellations of the angular perturbations becomes evident.
One can formulate  the statement:

(i) if
\be
e^{\frac{1}{2i}\int_0^T dtdt'
\Theta (t'-t) \hat{\phi}(t)\hat{e}_{\phi} (t')} e^{-iV_T (x_c ,e_c )}=
e^{-iV_T (x_c ,e_c )}|_{e_{\phi}=\phi =0} + dF(\phi_0 )/d\phi_0 ,
\l{49}
\ee
and

(ii) if
\be
F(\phi_0 +2\pi )=F(\phi_0 ),
\l{50}
\ee
then we easily find:
\ba
R(E)=2\pi\int^{2\pi}_{0} \frac{d\phi_0}{\Omega (E)}
\int^{\infty}_{0} dT dI_0 e^{\frac{1}{2i}\int_0^T dtdt'
\Theta (t'-t) (\hat{I}(t)\hat{e}_I (t')}
e^{S(x_c +e\partial x_c /\partial \phi_0 )-
S(x_c -e\partial x_c /\partial \phi_0 )}.
\l{51}
\ea
For the $(\lambda x^4)_1$-model
\be
S(x_c +e\partial x_c /\partial \phi_0 )-
S(x_c -e\partial x_c /\partial \phi_0 )=
S_0 (x_c)-
2\lambda \int^{T}_{0} dt x_c (t) \{e\partial x_c /\partial \phi_0 \}^3 ,
\l{53}
\ee
where \C{yp}
\be
S_0 (x_c)=
\oint_T dt (\frac{1}{2}\dot{x}_c^2 -
\frac{\omega_{0}^{2}}{2}x_c^2-\frac{\lambda}{4}x_c^4)
\l{54}
\ee
is the closed time-path action and
\be
x_c (t)=x_c (I(E)+I(t)-I(T), \phi_0 +\tilde{\Omega}t).
\l{52}
\ee
(here $I(t)$ and $I(T)$ are the auxiliary variables). In this
case the  problem is quasiclassical over the angular degrees of
freedom.

The condition (\ref{50}) requires that the  classical trajectory $x_c$, with
all derivatives over $I_0$, $\phi_0$, is the periodic function. In the
considered case of $(\lambda x^4)_1$-model $x_c$ is periodic function
with period $1/\Omega$ \C {tb}, see (\ref{ii}). Therefore, we can
concentrate our attention on the condition (\ref{49}) only.

Expanding $F(\phi_0)$ over $\lambda$:
\be
F(\phi_0)= \lambda F_1 (\phi_0)+ \lambda^2 F_2 (\phi_0)+...
\l{55}
\ee
we find frof (\ref{48}) that
\ba
\frac{d}{d\phi_0}F_1 (\phi_0)=
\int^{T}_{0}\prod^{3}_{k=1}dt'_k \hat{\phi}(t'_k )(-\frac{6}{(2i)^3})
\int^{T}_{0}dt \prod^{3}_{k=1}\Theta (t-t'_k )
x_c (t)(\partial x_c /\partial I_0 )^3 e^{iS_0 (x_c )}=
\n \\=
\int^{T}_{0} dt' \hat{\phi}(t')\{(-\frac{6}{(2i)^3})
\int^{T}_{0}dt \Theta (t-t')\prod^{2}_{k=1}(\Theta (t-t'_k ) \hat{\phi}(t'_k ))
x_c (t)(\partial x_c /\partial I_0 )^3 e^{iS_0 (x_c )}\} \equiv
\n \\ \equiv
\int^{T}_{0} dt' \hat{\phi}(t')B_1 (\phi ).
\l{56}
\ea
This example shows that the sum over all powers of $\lambda$ can be
written in the form:
\be
\frac{d}{d\phi_0}F(\phi_0)=
\int^{T}_{0} dt' \hat{\phi}(t')B(\phi ),
\l{57}
\ee
where, using the definition (\ref{41}),
\be
B(\phi )=\int^{T}_{0}dt \tilde{B}(\phi_0 +\phi (t)).
\l{58}
\ee
Therefore,
\be
\hat{\phi}(t')B(\phi )=\frac{d}{d\phi_0}
\int^{T}_{0} dt \delta (t-t') \tilde{B}(\phi_0 +\phi (t))
\l{59}
\ee
is coincide with the total derivative over initial phase $\phi_0$,
and
\be
F(\phi_0)=\tilde{B}(\phi_0 +\phi (t))|_{\phi=0}.
\l{60}
\ee
This result ends the consideration. It assumes that the
expansion over interaction constant $\lambda$ exist. Indeed, it is
known \C {tich} that the perturbation series for $(\lambda x^4)_1$-model
with $\lambda >0$ is  convergent in Borel sense.

\section{Conclusion}
\setcounter{equation}{0}

1. It was shown that the real-time quantum problem can be qusiclassical over
the part of the degrees of freedom and quantum over another ones.
Following to the result of this paper one may introduce the
(probably naive) interpretation of the quantum systems integrability
(we suppose that the classical system is integrable and can be
mapped on the compact hypersurface in the phase space \C{arn}): the quantum
system is strictly integrable in result of cancelation of all
quantum degrees of freedom. The mechanism of cancelation of the quantum
corrections is varied from case to case.

For some problems (as the rigid rotator, or the Pocshle-Teller)
the cancelation of quantum angular degrees of the freedom is enough since
they carry only the angular ones. In an another case (as in the
Coloumb problem, or in the one-dimensional models) the problem may
be partly integrable since the quantum fluctuations of action degrees
of freedom just survive. Theirs absence in the Coloumb problem needs
special discussion (one must take into account the dynamical (hidden)
symmetry of Coloumb problem \C{col}; to be  published).

The transformation to the action-angle variables maps the $N$-dimensional
Lagrange problem on the $2N$-dimensional phase-space torus. If the winding
number on this hypertorus is a constant (i.e. the topological charge
is conserved) one can expect the same cancellations. This is important
for the field-theoretical problems (for instance, for $sine$-Gordone
model \C{sg}).

2. In the classical mechanics the following approximated method of
calculations is used \C{arn}. The canonical equations of motion:
\be
\dot{I}=a(I,\phi),\;\;\;\;\;\dot{\phi}=b(I,\phi)
\l{i}
\ee
are changed on the averaged equations:
\be
\dot{J}=\frac{1}{2\pi}\int^{2\pi}_{0} d\phi a(J,\phi),\;\;\;\;\;
\dot{\phi}=b(J,\phi),
\l{ii}
\ee
It is possible if the periodic oscillations can be extracted
from the systematic evolution of the degrees of freedoms.

In our case
\be
a(I,\phi)=j\partial x_c /\partial \phi,\;\;\;\;\;
b(I,\phi)=\Omega (I)- j\partial x_c /\partial I.
\l{iii}
\ee
Inserting this definitions into (\ref{ii}) we find evidently wrong
result since in this approximation the problem looks like pure quasiclassical
for the case of periodic motion:
\be
\dot{J}=0,\;\;\;\;\;\dot{\phi}=\Omega(J).
\l{iv}
\ee
The result of this paper was used here. This shows that the procedure of
extraction of the periodic oscillations from the systematic evolution is
not trivial and this method should be used carefully in the quantum theories.
(This approximation of dynamics is ``good" on the time intervals
$\sim 1/|a|$ \C{arn}.)

\vspace{0.2in}
{\large \bf Acknowledgement}
\vspace{0.2in}

I would like to thank A.Ushveridze and I.Paziashvili for stimulating
discussions.

\end{document}